\documentclass{aa}   
\usepackage{graphicx}
\usepackage[varg]{txfonts}
\begin{document}

\title{Quasi integral of motion for axisymmetric potentials}


\author{O. Bienaym\'e\inst{1}
\and A.C. Robin \inst{2}
\and B. Famaey \inst{1}
}
 
\institute{Observatoire astronomique de Strasbourg, Universit\'e de Strasbourg, CNRS, UMR 7550, 11 rue de l'Universit\'e, F-67000 Strasbourg, France 
        \and Institut Utinam, CNRS UMR 6213, Universit\'e de Franche-Comt\'e, OSU THETA Franche-Comt\'e-Bourgogne, Observatoire de Besan\c con, BP 1615, 25010 Besan\c con Cedex, France}

   \date{Accepted, July 2015}

 
  \abstract
   {We present an  estimate of the third integral of motion for axisymmetric three-dimensional potentials. This estimate is based on a St\"{a}ckel approximation and is explicitly written as a function of the potential. We tested this scheme for the Besan\c con Galactic model and two other disc-halo models and find that orbits of disc stars have an accurately conserved third quasi integral. 
The accuracy  ranges from  of 0.1\% to 1\% for heights varying from $z$ = 0~kpc to $z=$ 6\,kpc and Galactocentric radii $R$ from 5 to 15\,kpc.
We also tested the usefulness of this quasi integral in analytic distribution functions of disc stellar populations: we show that the distribution function remains approximately stationary and that it allows to recover the potential and forces by applying Jeans equations to its moments.}


   \maketitle
%



\section{ Introduction}

Many approaches exist for the galactic modelling of stellar dynamics. With the availability of a large amount of new accurate data for the velocity and position of stars in the extended solar neighbourhood, the development of the most precise tools for analysing the kinematics of stars is becoming inescapable.

Numerical modelling techniques are the most direct ones, such as for the Scharwschild method, the made-to-measure method  \citep{Syer96}, or of course ab initio N-body and hydrodynamical simulations \citep{Renaud13}. Spectral analysis \citep{Papaphilippou98,Valluri98} also offers useful numerical tools for recognising and identifying resonances. Numerical resolution of the collisionless Boltzmann equation is another  direct way to model, but it is still limited by available numerical resources \citep{Yoshikawa13, Colombi15}.

Analytical techniques are more explicit but imply analytical approximations. In the case of axisymmetric galactic potentials, it is known that orbits are generally constrained by an effective third integral of motion in addition to the energy and angular momentum. From the Jeans theorem, therefore, equilibrium models of axisymmetric galaxies should be represented by distribution functions that only depend on these three integrals. The modelling of stellar distribution functions can thus be achieved if the effective third integral can be approximated analytically or numerically. 

A long list of works in this direction already exists, and very useful bibliographies may be found in \citet{deZeeuw85a}, \cite{deZeeuw85b}, \cite{deBruyne00}, \cite{Binney12}, and \cite{Sanders14}, among others. We concentrate here on the use of St\"ackel potentials that have been shown in many cases  to be efficient for modelling different families of orbits. Published works on this subject present a wide variety of approaches, and we  can distinguish between the local and global approaches. The local modelling  consists  in locally fitting the true potential with a St\"ackel potential. This allows  nearly exact modelling of orbits and  integrals of motion  to be obtained in the immediate neighbourhood of the position considered.

Global modelling, either of the potential or of the orbits over a predefined phase-space volume, has opened the way to many different technicalities over the past 50 years. In each case the goal was to use a St\"ackel third integral as an approximate of the third quasi integral of the fitted potential, as
in \cite{Wayman59}, \cite{vandeHulst62},  \cite{Ollongren62}, 
\cite{Hori62}, and more recently, \cite{Manabe79}, \cite{Batsleer94}, or \cite{Famaey03}.

A recent novel application of the St\"ackel potential approximations has been to  model the local stellar kinematics and the gravitational potential in the solar neighbourhood at large $z$ distances  up to 1-2 kpc from the galactic plane \citep{Bienayme14,Piffl14}. These recent studies put  strict constraints on the vertical variation in the gravitational potential and in the local density of the dark matter halo. Recently, \cite{Sanders14} have proposed a global St\"ackel fitting of 3D potentials, and they  give  algebraic expressions that  allow recovery of the integrals of motion, expressed as action variables, and they present numerical applications.

In the present paper, we proceed to a St\"ackel potential fitting by using  a simple expression for the integral of motion that explicitly depends on the  potential.  Our study has similarities with the works recently published by \cite{Binney12} or by \cite{Sanders14}, proposing different formulations of a third integral. Although there are also advantages in working with action integrals, the present approach is, however, much simpler and more straightforward to apply.

The  paper proceeds as follows. In Section 2, we   give a new  expression for the third integral, and its derivation is described in the Appendix. In Section 3 we  examine,  for three potentials,  the constancy of this third integral along orbits. Section 4 shows, in the case of the Besan\c con Galactic model, how moments of distribution functions based on this third integral are in accordance with Jeans equations. Section 5 finally considers the application to the collisionless Boltzmann equation.
  
\section{Quasi integral of motion}

We let $V(R,z)$ be an  axisymmetric gravitational potential. Besides the energy $E$ and the vertical component of the angular momentum $L_z$, which are integrals of motion, we define here, as an approximate third integral of  motion,  

\begin{equation}
\label{eq1}
I_s= \Psi(R,z)
-\frac{1}{2} \frac{L^2-L_z^2}{z_0^2} -\frac{1}{2}v_z^2
\end{equation}
where $L$ and $L_z$ are the total and vertical angular momenta, $z_0$ a fixed parameter at fixed $E$ and $L_z$, $v_z$ the vertical velocity, and with
\begin{equation}
\label{eq2}
\Psi(R,z)=-
\left[ V (R,z) 
-   V(\sqrt{\lambda},0 )\right] \, \frac{(\lambda +z_0^2)}
{z_0^2}\, ,
\end{equation}
and
\begin{eqnarray}
\label{eq3}
 \lambda  = & \frac{1}{2}(R^2+z^2-z_0^2)
+\frac{1}{2}\sqrt{\,(R^2+z^2-z_0^2)^2+4\,R^2\, z_0^2} \, .
\end{eqnarray}

This approximate integral $I_s$ is  an exact integral in the case of St\"ackel potentials expressed in an ellipsoidal coordinate system of focus $z_0$, which is thus a known quantity. Its derivation is explained in the Appendix. 

The only free quantity for determining $I_s$, when $V(R,z)$ is known, is the parameter $z_0$. Here, it will be numerically adjusted by minimizing $\sigma_{Is}$ the dispersion of the quasi integral $I_s$ along all the orbits with the same energy and vertical angular momentum. 
Thus $z_0$ will itself be a function of the two integrals $E$ and $L_z$.

In a preliminary analysis, which was not developed further, we used the generic differential equation  of St\"ackel potentials that can be rewritten to determine $z_0$
locally as a function of the second derivatives of the potential at any positions 
\citep[see Eq.~10 in][]{Bienayme09, deZeeuw85a}.
However,  we find out that  such a procedure gives us significantly fewer accurate results than fitting a single $z_0$ for a given orbit or family of orbits. Here, we  look for the $z_0$ value that minimizes the dispersion of $I_s$ along each orbit of a family of orbits with the same $E$ and $L_z$. We note, however, that for any ($E$,$L_z$)-family of orbits and with $z_{shell}$ being the maximum $z$-extent of the shell orbit, the value of $z_0$ that gives the best fit is  close to the value of $z_0$ obtained by applying Eq. 10 of \cite{Bienayme09} at position $(R_c, z_{shell}/2)$ (with $R_c(L_z)$  the  radius of the circular orbit with angular momentum $L_z$). 

The degree of the approximation of the quasi integral $I_s$ can be controlled in several ways: (1) inspection of the surfaces of section, (2) conservation of orbital weights and of  the spatial density, and (3) conservation of $I_s$ along the orbits to validate the labelling of orbits. The conservation of orbital weights and the conservation of $I_s$ are the criteria that can be best expressed numerically.

In the following sections, we test the degree of conservation of this quasi integral along orbits in the cases of  three non-St\"ackel potentials. We also determine its  efficiency to build stationary stellar disc distribution functions and to measure the potential from the Jeans equations. \\

\section{Testing the conservation of the  integral along orbits}

We examined the  conservation of the quasi-integral of motion along orbits of stars corresponding to disc stellar populations. For these disc components, stellar motions have restricted oscillations in the radial and vertical directions. For the smallest oscillations, the phase-space domain explored by stars is the closest to a St\"ackel potential, and the quasi integral tends to be constant. 

We considered stars with  identical angular momentum $L_z$ (with $E_c$ the energy of the corresponding circular orbit) and  identical energy
$E$  (we define $\Delta E=E-E_c$). For each star, we computed the mean value of $I_s$ and its dispersion $\sigma_{Is}$  from $\sim$10~000 steps along the orbit, 
and we determined the value of $z_0$ that minimizes $\sigma_{Is}$.
To ensure that $\sigma_s$ is determined sufficiently well, the total time integration for each orbit is about 300 dynamical times (i.e. 300 galactic rotations). 
The numerical integration is a Runge-Kutta-Fehlberg of order 7(8) \citep{Fehlberg68}. We find that  stellar orbits with the same values of $L_z$ and $\Delta E$  all have  $\sigma_{Is}$  minimum for similar values of $z_0$.
For different energies and angular momenta, the adjusted $z_0$ will be different.

To allow an easier presentation of our results, we   normalized  $I_s$ as \\
\begin{equation}
\label{eq4}
I_3=-\frac{I_s}{\Delta E} \left(1+\frac{R_{c}^2(L_z)}{z_0^2} \right)^{-1} \, , 
\end{equation}
still an integral of motion.  $I_3$=0 corresponds to orbits confined within the galactic plane, while $I_3$$\sim$1 corresponds to  shell orbits with minimum radial variations when they cross the plane $z=0$.

In the following sections, we consider three potentials: a  logarithmic potential with a disc component, a flattened logarithmic potential, and a potential  similar to the Besan\c con Galactic model potential. In all the three cases, the constancy of the third integral is  satisfied  at  0.2 to 0.5 per cent for orbits corresponding to thin or thick stars and to a 1 to 2 per cent for orbits with  larger vertical oscillations.

\begin{table}[htdp]
\caption{  Disc-halo logarithmic potential. Maximum and median values of the quasi-integral dispersion  $\sigma_{I3}$ (i.e. the relative error) for orbits with different energies $\Delta E$:  best fit of $z_0$, maximal height $z_{max}$ above the galactic plane for the shell orbits.}
\begin{center}

\begin{tabular}{ c c c c c c c c c c }
\hline\hline
$\Delta$ E      &        $z_0$  &       $z_{max}$       &       $ \sigma_{I3}$ maximum  &      $ \sigma_{I3}$ median  \\
\hline
0.05            &       3.3             &       2.4                     &       0.008                                           &       0.004 \\
0.1                     &       3.3             &       3.7                     &       0.006                                           &       0.003 \\
0.2                     &       3.3             &       6.1                     &       0.021                                           &       0.005 \\
0.5                     &       3.3             &       12.9            &       0.012                                           &       0.006 \\
\hline
\end{tabular}
\end{center}
\label{default1}
\end{table}%

\subsection{Logarithmic potential of Myamoto-Nagai type}

The logarithmic potential of Myamoto-Nagai type 
\begin{equation}
\label{eq5}
 \Phi=  v_\infty^2 \log \left(  R^2 + \left(a+\sqrt{b^2+z^2} \right)^2  \right)^{1/2} 
\end{equation}
has a positive density \citep{Zotos11}. It  has a spherical halo and a disc component (with $a+b$ the halo core radius and $b$ the disc thickness). This potential presents analogy  with the family of density-potential  pairs of logarithmic Myamoto-Nagai type given by \cite{Bienayme09}.

We consider the orbits with the angular momentum $L_z$=8.5 ($v_\infty$=1, $a$=0.5, $b$=0.2,  $R_{c}$ =8.5) and 
with $E=E_c+\Delta E$. The initial conditions of the computed orbits are $v_R=0$, with $R_{initial}$ equally spaced. These initial conditions do not include resonant orbits that do not cross the galactic plane  perpendicularly.

Table 1 summarizes the results. For each energy $\Delta E$, the adjusted $z_0$ that minimizes  $\sigma_{I3}$ is given. The  maximum vertical extension $z_{max}$ corresponds to an orbit with $I_3$ close to 1. The dispersions of $I_3$ always remain small, the mean dispersion being smaller than one per cent. However, for some resonant orbits, the dispersion is larger, of the order of 2 per cent. We also note that at large energies corresponding  to halo stars the variation of $I_3$ still remains  small. This must be  linked to the spherical shape of the potential at large $z$, easily modelled by a St\"ackel potential. 

In the next section we consider a different requirement using a flattened potential. 
\\

\begin{figure}[!htbp]
\begin{center}
\resizebox{8.5cm}{!}{\rotatebox{0}{\includegraphics{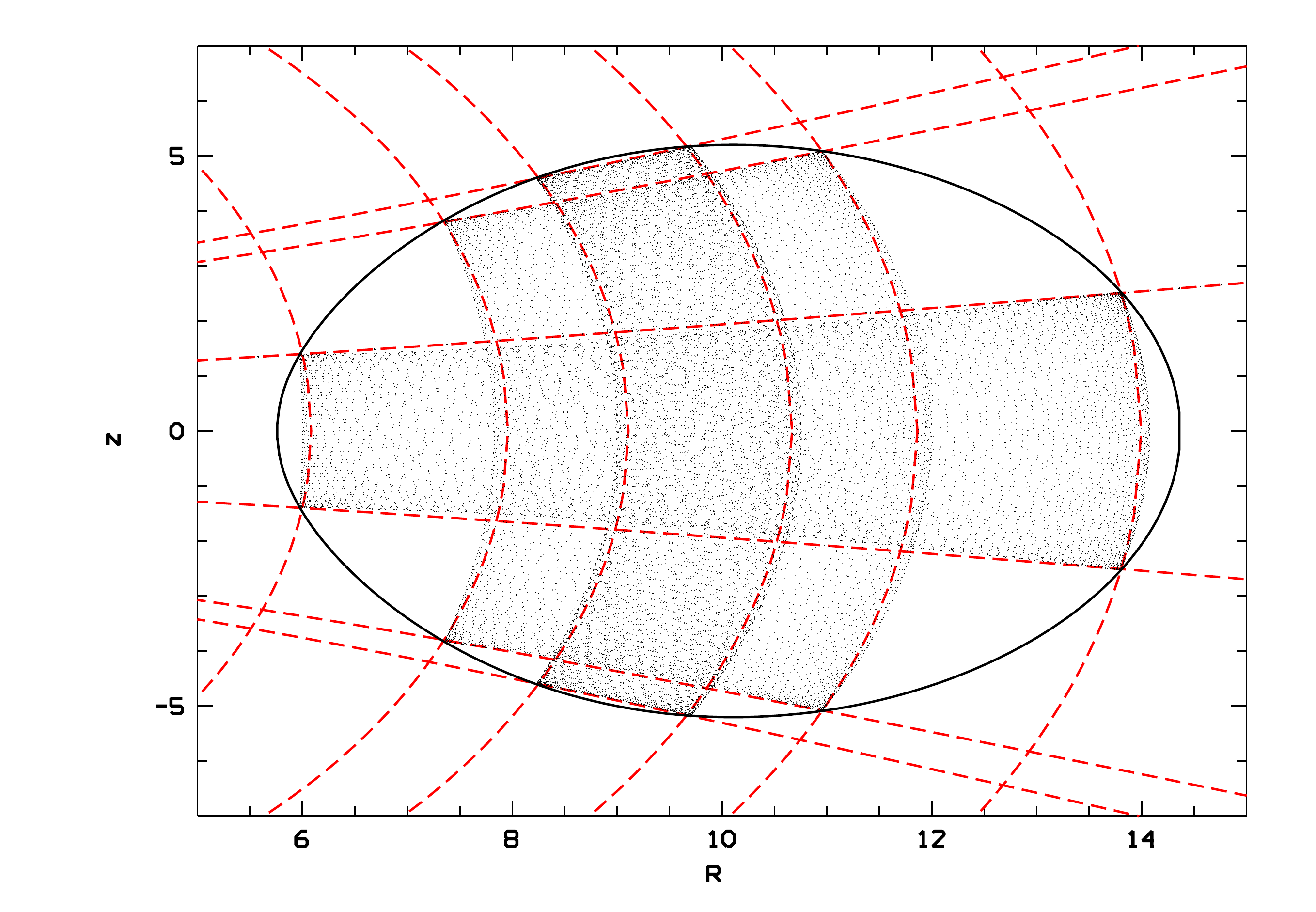}}}
\resizebox{8.5cm}{!}{\rotatebox{-90}{\includegraphics{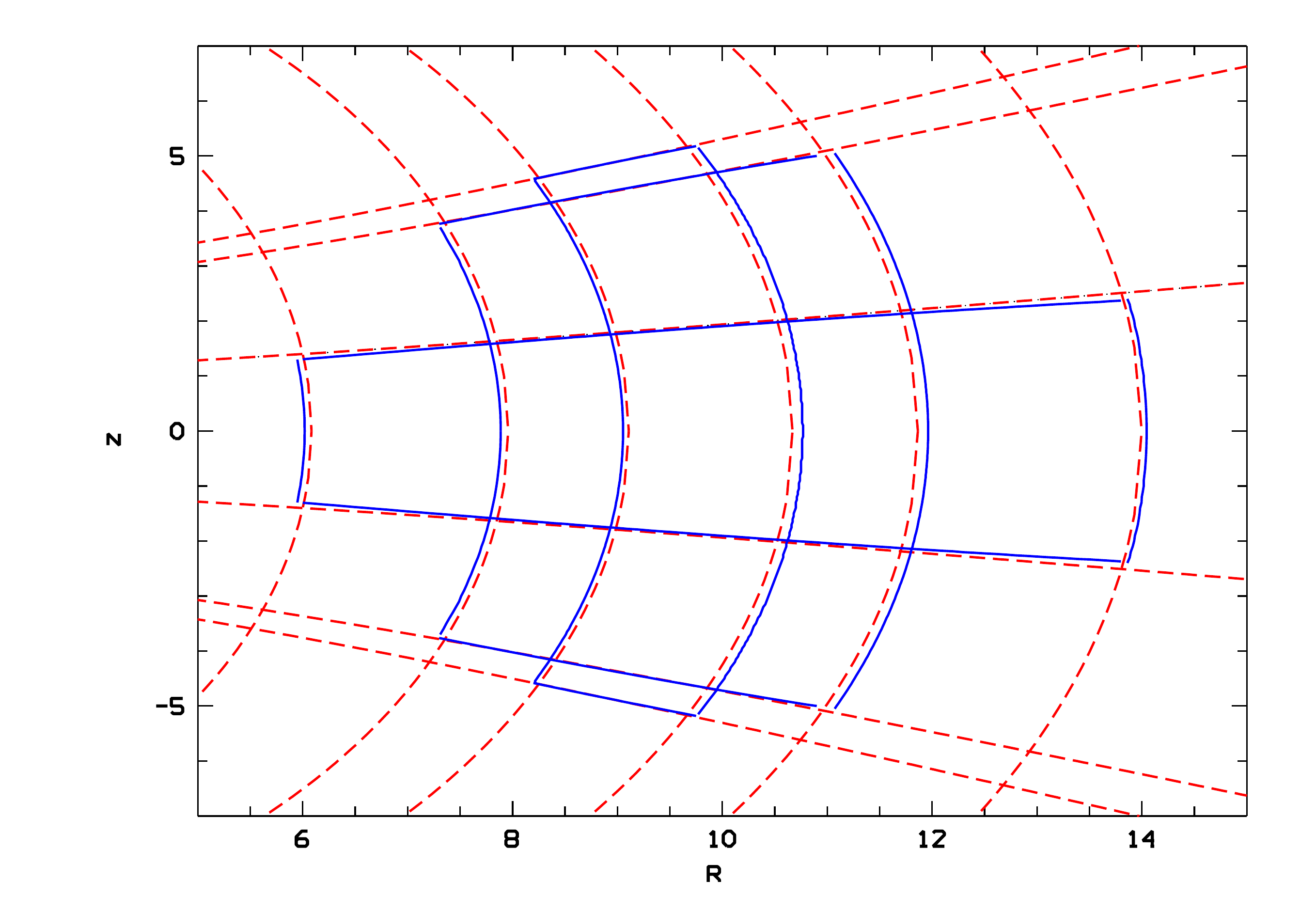}}}
\resizebox{8.5cm}{!}{\rotatebox{-90}{\includegraphics{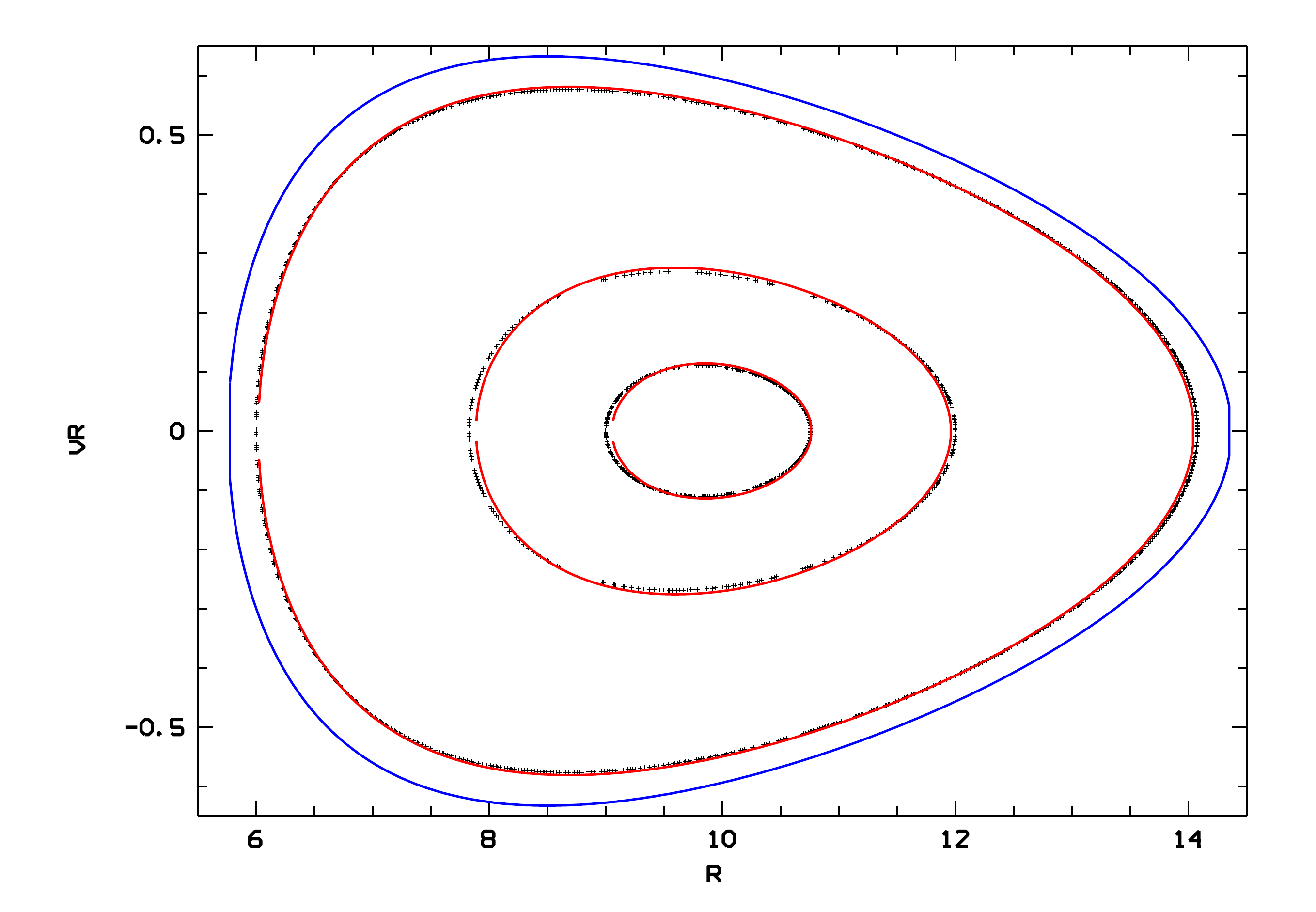}}}
\end{center}
  \caption{
  Top panel: Meridional projection for three orbits within the logarithmic potential (Sect. 3.2) with $L_z$=8.5 and $\Delta E$=0.2. Zero velocity curve and some coordinate curves of the elliptic coordinates (with $z_0$=5.9) are drawn.
  Middle panel: Analytical determination of the envelopes of the orbits.
  Bottom panel: Surfaces of section for the same three orbits. Black crosses: numerically computed orbits.  Red lines: the corresponding  sections obtained from  Eqs~\ref{eq1}-\ref{eq3}. Blue line: surface of section of the orbit confined in the mid-plane (i.e. $v_z$=0).  }
    \label{fig1}
\end{figure}

\begin{figure*}[!htbp]
\begin{center}
\resizebox{8.5cm}{!}{\rotatebox{-90}{\includegraphics{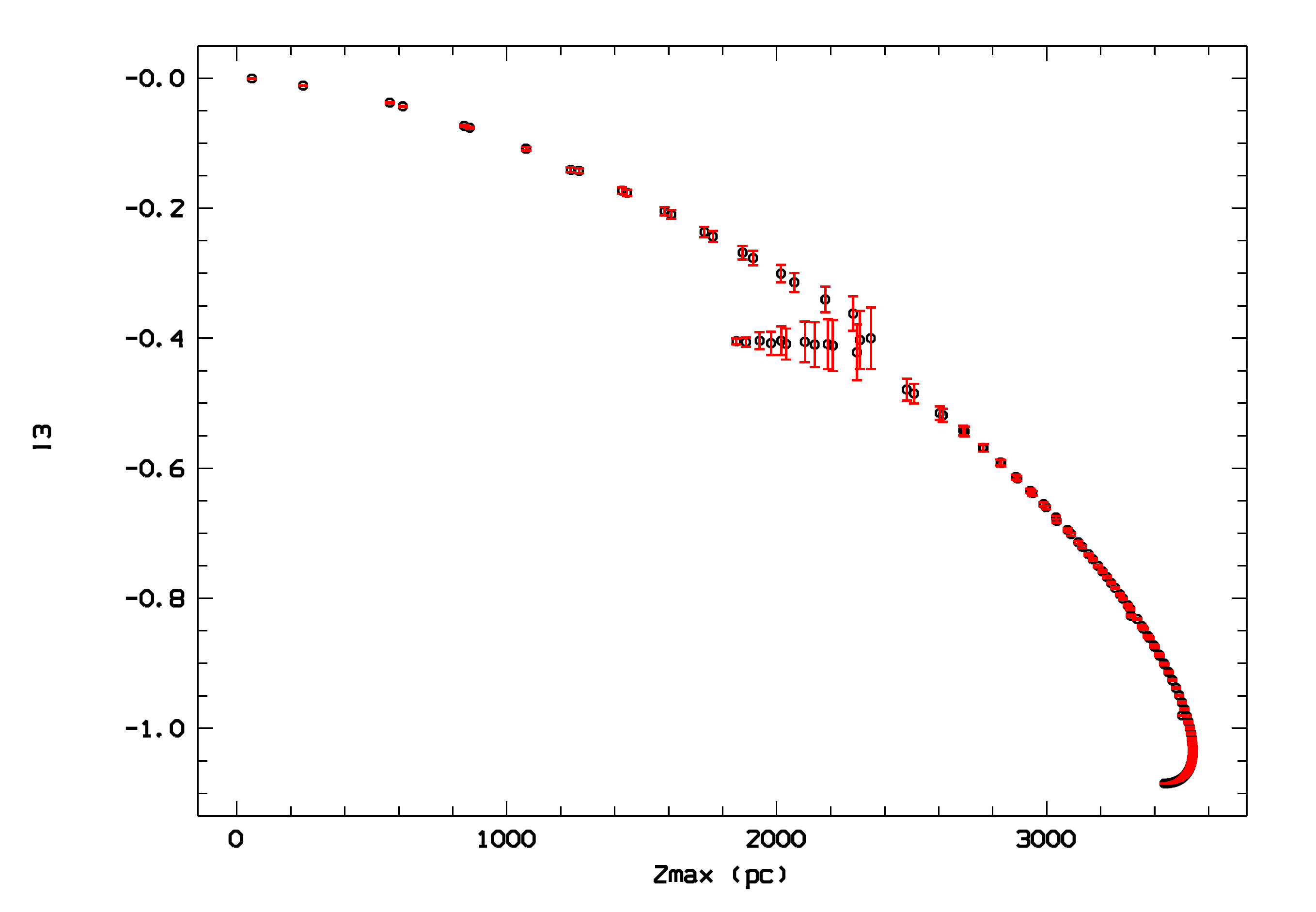}}}
\resizebox{8.5cm}{!}{\rotatebox{-90}{\includegraphics{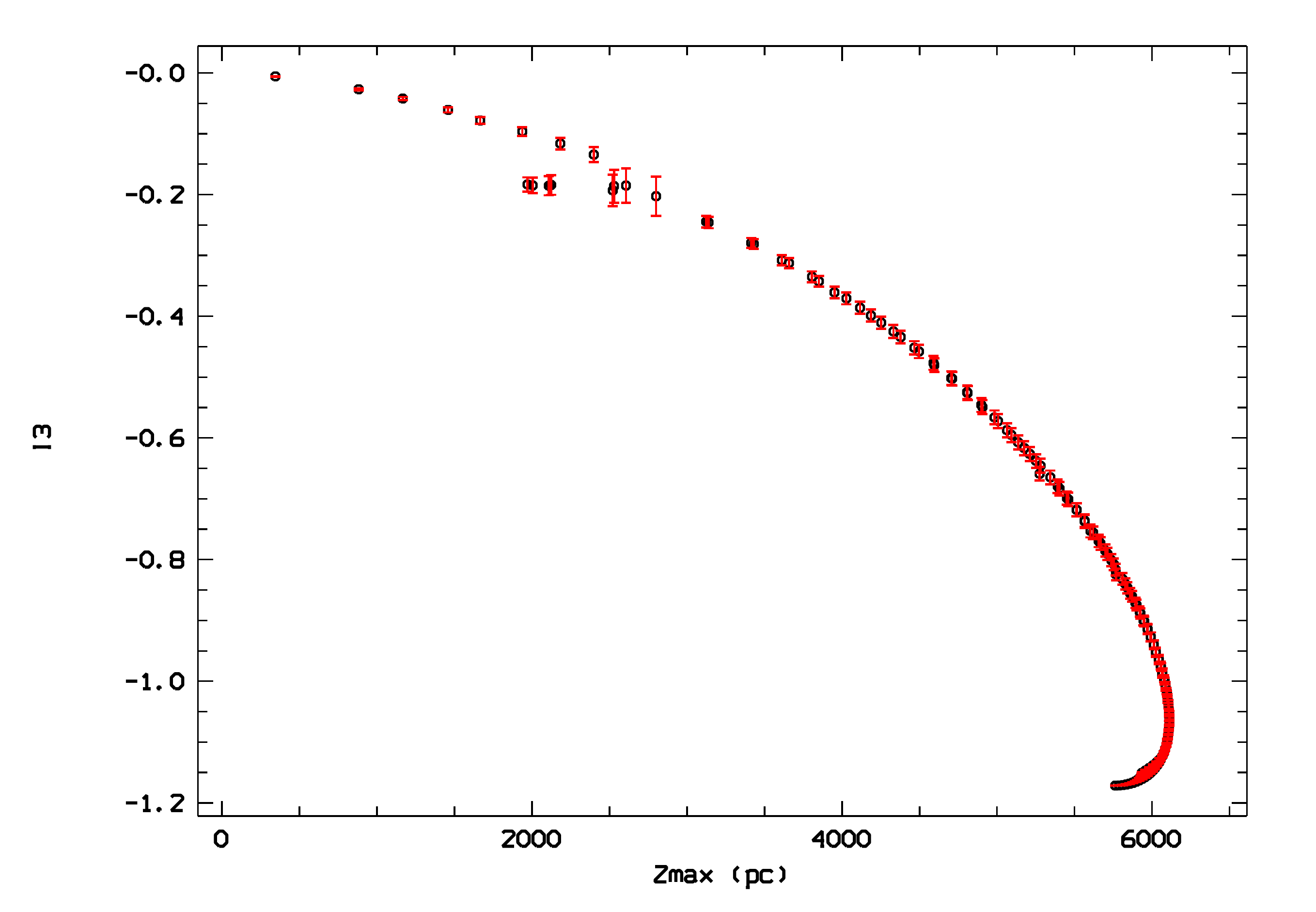}}}
\end{center}
  \caption{Left: quasi integral $I_3$ versus $z_{max}$ for stars with $\Delta E=6400$ and $R_c(L_z)$=8500\,pc for the BGM potential. Errors bars are $\sigma_{I3}$.
 Right: idem with $\Delta E=12800$.
    }
    \label{fig2}
\end{figure*}


\begin{figure*}[!htbp]
\begin{center}
\resizebox{8.5cm}{!}{\rotatebox{-90}{\includegraphics{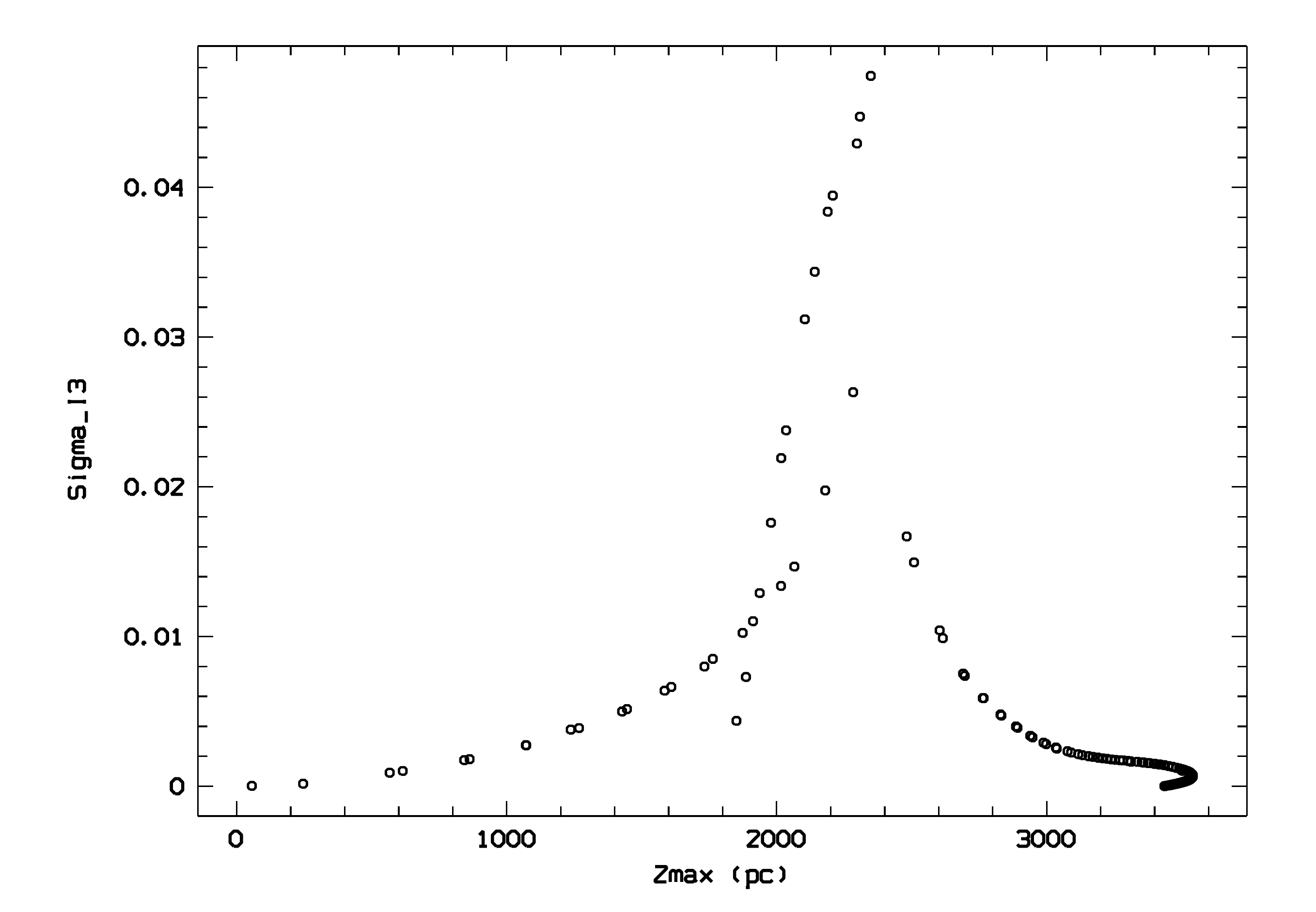}}}
\resizebox{8.5cm}{!}{\rotatebox{-90}{\includegraphics{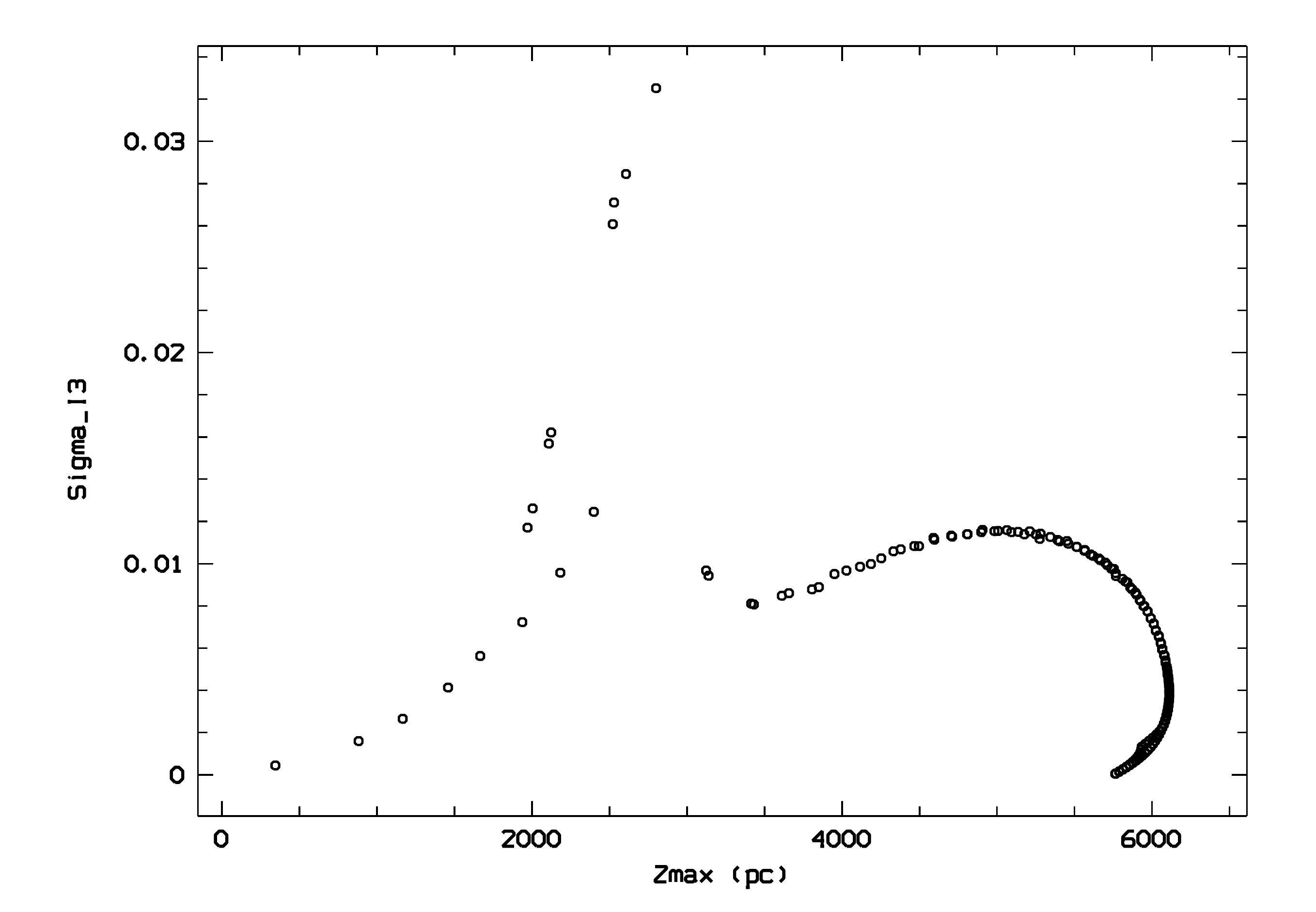}}}
\end{center}
  \caption{ Left: Dispersion $\sigma_{I3}$ of the integral $I_3$ along orbits versus $z_{max}$ for stars with $\Delta E=6400$ and $R_c(L_z)$=8500 for the BGM potential. 
  Right: the same with $\Delta E=12800$.
    }
    \label{fig3}
\end{figure*}

\subsection{Logarithmic potential}

To examine the impact of the flattening of the dark halo on the dispersions $\sigma_{I3}$, we consider the traditional logarithmic potential \cite{Richstone80}:
\begin{equation}
\label{eq6}
 \Phi= \frac{v_c^2}{2} \log \left( R^2+ z^2/q^2 \right) 
.\end{equation}
We set $v_c$=1 and $q=0.8$ to correspond to a flattening of the isodensity $q_\rho=0.65$. 

We computed orbits with $L_z=8.5$ ($R_{c}=8.5$, $E_c=1$). Table 2 summarizes the results. Surfaces of section for $L_z=1$, $\Delta E=0.05,$ and 0.5 are shown in Fig. 9 of  \cite{Bienayme13}, and we note the lack of resonant orbits. This  may explain the excellent conservation of $I_3$ at a  level of 0.2 per cent for  stellar disc orbits. Orbits with larger extension up to  $z_{max}$=9 ($E$=0.4) have $I_3$ conserved at 2 per cent.

The inspection (see Fig. \ref{fig1}) of the surface of section and of the meridional projection of three orbits ($\Delta E$=0.2 and $L_z$=8.5) allows visualization of the relative agreement between the nearly exact  numerical orbits and the contours obtained from the approximated third integral. For the meridional projection, the red dashed lines are ellipsoids and hyperboloids from the ellipsoidal coordinates with $z_0$=5.9 (lines are drawn from the corner of the orbit envelopes).
The blue continuous  lines in Figure~\ref{fig1} (middle panel) are the envelopes that are determined semi-analytically from  $E$, $L_z$ and the quasi  integral $I_3$ that are explicitly known: we determine surfaces of section at various fixed $R$ (resp. fixed $z$) and find the $z$ (resp. $R$) extension limits of the orbit from the condition $\partial I_3/\partial v_r$=0 (resp. $\partial I_3/\partial v_z$=0). We note small differences between the envelopes of numerical orbits (top panel) and semi-analytical envelopes (middle panel). We also note that  the analytical hypersurfaces defined from $E$, $I_3$ do not have  the topology of a torus everywhere. 
The bottom panel of Figure~\ref{fig1}  shows a section of the phase space ($z$=0) for the same three orbits from the numerically computed   orbits and from the analytical section given by Eq \ref{eq1}. If $v_c$=200 km/s, the maximum $v_R$ of these three orbits is $\sim$120, 50, and 20 km/s.

\begin{table}[htdp]
\caption{Logarithmic potential (see legend of Table 1).}
\begin{center}
\begin{tabular}{c c c c c c c c c c }
\hline\hline
$\Delta$ E              &                $z_0$     &      $z_{max}$             & $ \sigma_{I3}$ maximum  &      $ \sigma_{I3}$ median\\
\hline
0.04    &       5.4     &       2.0     &       0.004   &       0.002\\
0.1             &       5.5     &       3.3     &       0.011   &       0.005\\
0.2             &       5.9     &       5.2     &       0.020   &       0.010\\
0.4             &       7.0     &       8.8     &       0.040   &       0.019\\
0.8             &       10.     &       17.5&   0.08    &       0.04\\
\hline
\end{tabular}
\end{center}
\label{default2}
\end{table}%

\subsection{The Besan\c con Galactic mass model}

 We apply our test to   the gravitational potential of the Besan\c con Galactic model (BGM) \citep{Bienayme87, Robin03, Czekaj14},  which is a more realistic one than the two previous  ones, although we limited our  analysis to its axisymmetric version. This model has a dynamical consistency at the solar Galactic radius  position in  the sense that the thickness of the stellar components are constrained by the potential and by the vertical velocity dispersions.  However, outside of the solar neighbourhood, the  vertical kinematics  and the thickness of stellar components  are constrained by observations and not by a dynamical consistency of the model. The motivation of this work is to improve the BGM dynamical consistency  by using stationary distribution functions to model the stellar kinematics. Besides the integrals $E$ and $L_z$, we therefore require an approximate third integral to describe the stellar kinematics.

We determined the gravitational potential of the Besan\c con Galactic model according to the characteristics of its  components  described in \cite{Robin03} and \cite{Czekaj14}. The dark matter component is represented by a spherical component.  The orbits with large vertical extensions have shapes  that are mainly determined by the spherical halo. Since the exact shape of the dark halo remains uncertain in practice and is a controversial question, we consider a less favourable situation using a flattened  dark matter component.

The dark matter halo  density used in the BGM is
\begin{equation}
\label{eq7}
\rho_{dm}\sim\frac{1}{1+(R/R_{core})^2}\, .
\end{equation}
We flatten the halo, replacing the quantity $R$ with $\sqrt{R^2+z^2/q^2}$ in the expression of the potential. With $R_{core}$=2700\,pc and $q=0.8$,  the  flattening of the density is $q_\rho=0.62.$ (For this value of $q$,  negative densities only occur in a region far from  our domain of interest.)

We also modify the point-mass bulge \citep{Bienayme87}  using the potential law 
$\Phi_B=1/\sqrt{R_{bulbe}^2+R^2+z^2}$
with $R_{bulbe}=2000$\,pc.

We summarize  the results  for orbits with $L_z$  corresponding to $R_c=$8500 pc in Table 3.  Results are listed by families of different energies $\Delta E$.
Figure \ref{fig2} shows $I_3$  versus  $z_{max}$, the maximum vertical extension of orbits in  cases $\Delta E$=6400 or 12800 and $R_c=$8500 pc.  Error bars are the dispersion of $I_3$ along each orbit. The dispersion of $I_3$ is zero for orbits confined in the mid-plane. Shell orbits have $I_3$ maximum and  $\sigma_{I3}$   close to a few $10^{-4}$ (Figure \ref{fig3}). This is unexpected since there is  no reason that the potential along the 'thin' orbits should have exactly the St\"ackel form. On the other hand, when $z_{max}$$\sim$2000\,pc, the dispersion of $I_3$ is maximum and increases sharply for all energies (see Figure~\ref{fig3}), also  corresponding  in Figure \ref{fig2} to a change of shape of the distribution of points. This is related to the presence  of (1,1) resonant orbits at this vertical height that are poorly modelled by our St\"ackel adjustment. For these orbits  the dispersion of the quasi integral is   less than 5 per cent.  The median dispersion of $I_3$ for orbits with $z_{max}$ smaller than 2\,kpc remains very small $\sim$0.002.

\begin{table}[htdp]
\caption{Besan\c con Galactic model with  $q$=0.8 and  $R_c$=8.5\,kpc (see legend of Table 1), where $w_{max}$ is the maximum vertical velocity.}
\begin{center}
\begin{tabular}{c c c c c c c c c c }
\hline\hline
$\Delta$ E & $w_{max}$ &  $z_0 $      &  $z_{max} $      &           $ \sigma_{I3}$ max. &        $ \sigma_{I3}$ med.      \\
\hline
km$^2$.s$^{-2}$          &       km.s$^{-1}$  &   kpc      &  kpc      &           &              \\
\hline

400     &       28.3    & 4.8   &       0.42    &       0.001   &       0.0004\\

1600&   56.6    & 4.4   &       1.0             &       0.002   &       0.0012\\

3200&   80              & 4.3   &       1.7             &       0.004   &       0.0025\\

6400&   113             & 4.5   &       3.5             &       0.047   &       0.0043\\

12800& 160              & 5.1   &       6.1             &       0.032   &       0.0070\\

\hline      
\end{tabular}
\end{center}
\label{default3}
\end{table}%

\begin{figure}[!htbp]
\begin{center}
\resizebox{8.5cm}{!}{\rotatebox{-90}{\includegraphics{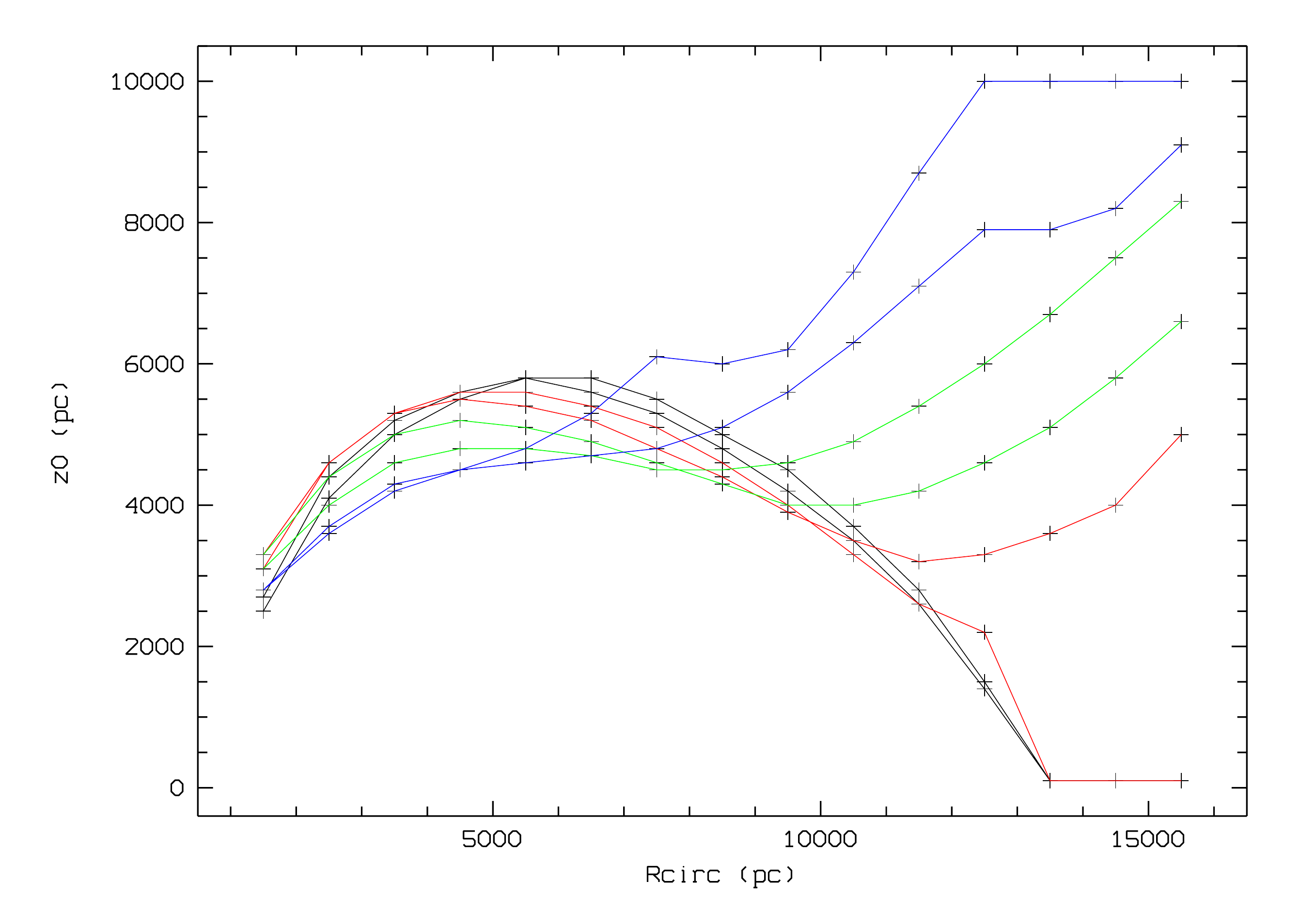}}}
\end{center}
  \caption{ BGM potential: $z_0$ best fit versus Rcirc(Lz) and $\Delta E$. Black lines $\Delta E$=100 and 200; 
  red lines: 400 and 800; green lines: 1600 and 3200; blue lines: 64000 and 128000.
    }
\label{fig4}
\end{figure}

We performed the St\"ackel adjustment for other galactic radii  $R_c$ from
1.5 to 15\,kpc. For each pair of values ($E=Ec+\Delta E$,  $L_z$), we determine the $z_0$ value that minimizes $\sigma_{I3}$.
Figure \ref{fig4}  plots the fitted $z_0$ (crosses) for  $\Delta E$=100,200,400...12800. Lines of constant $\Delta E$ are plotted.  We obtain a tabulation of $z_0$ that gives us a function of two integrals of motion $z_0(E,L_z)$ that can be used to build a distribution function for stellar populations.

\begin{figure}[!htbp]
\begin{center}
\resizebox{8.5cm}{!}{\rotatebox{-90}{\includegraphics{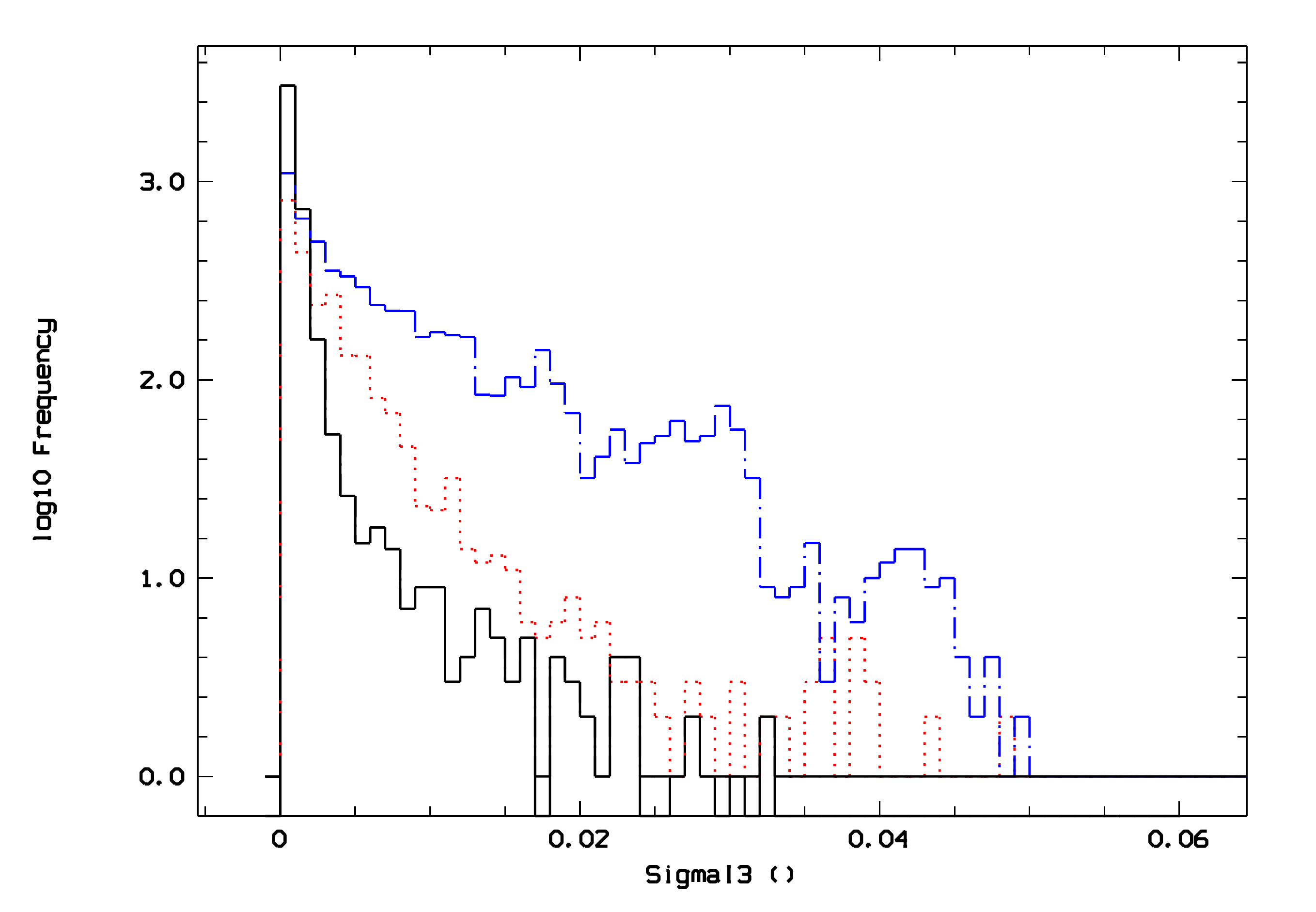}}}
\end{center}
  \caption{  Histograms of $I_3$ dispersion along orbits for $z_{max}$ intervals delimited by  0, 1000\,pc, 2000\,pc, and beyond (respectively black,  dotted red, dashed blue lines). For each of the three $z_{max}$ intervals, the medians of $\sigma_{I3}$  are respectively $5.\,10^{-4}$, $1.5\,10^{-3}$, and $5.\,10^{-3}$. 
    }
    \label{fig5}
\end{figure}

The median  dispersions of $I_3$  remain low except at  galactic radii smaller than 3\,kc
where many resonant orbits dominate the phase space.
Figure \ref{fig5}  plots the histograms of $\sigma_{I3}$ for $R_c$ from 3.5 to 15.5\,kpc. 

\section{Distribution function and Jeans equations}

%
%
\begin{figure*}[!htbp]
\begin{center} 
\resizebox{8.5cm}{!}{\rotatebox{-90}{\includegraphics{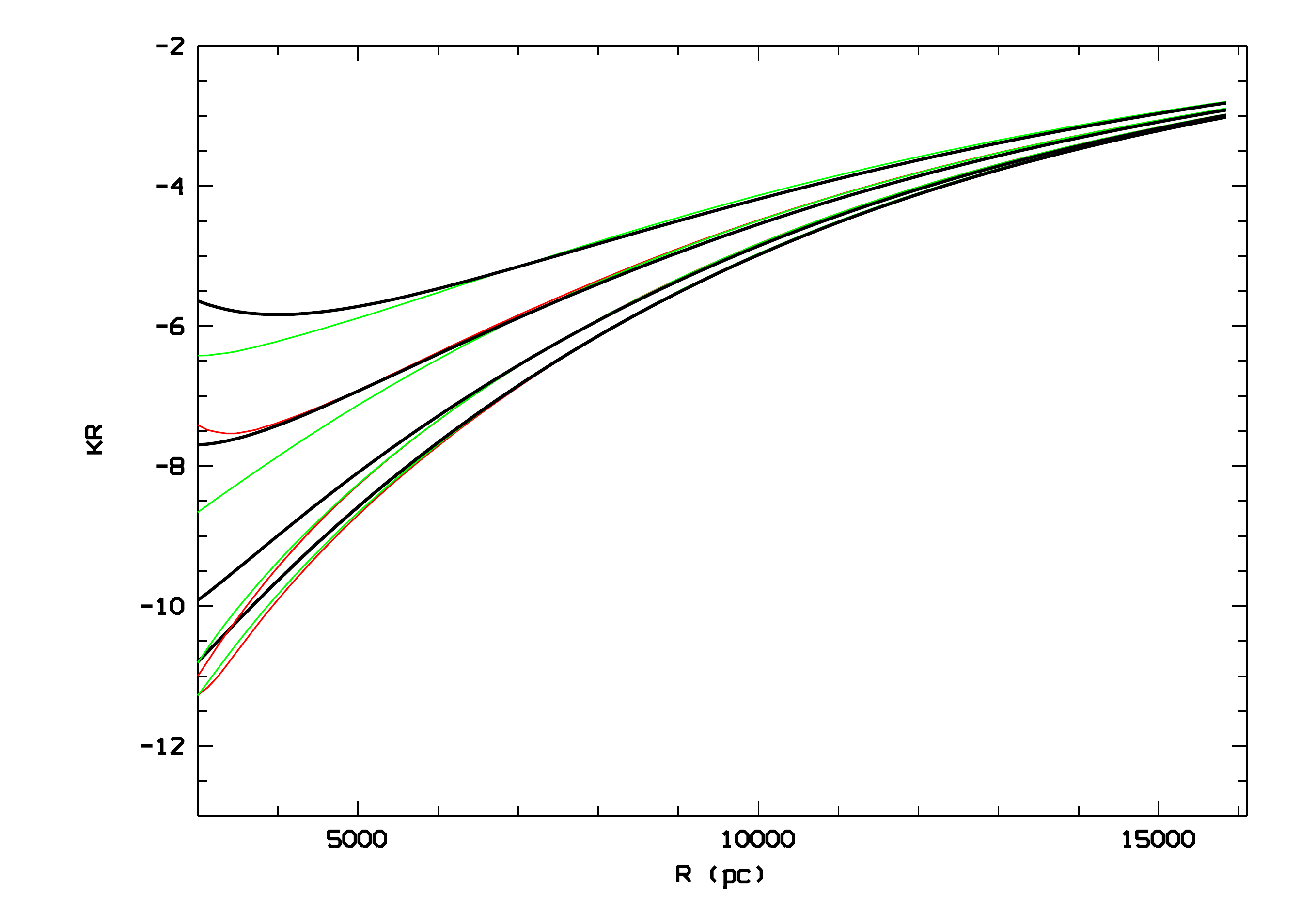}}}
\resizebox{8.5cm}{!}{\rotatebox{-90}{\includegraphics{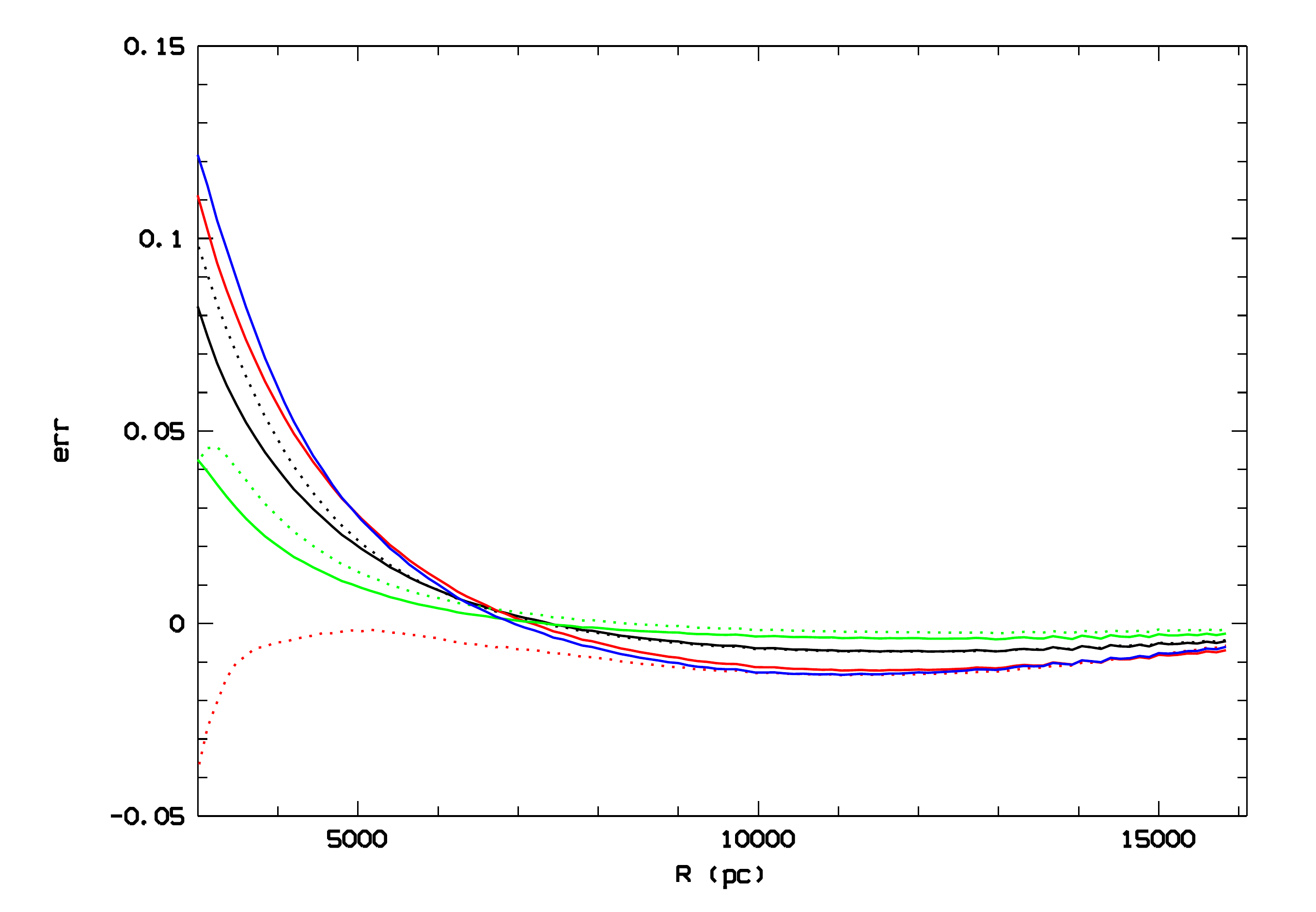}}}
\end{center}
  \caption{ (Left) Dark lines: radial forces $K_R$ of the BGM at several $z$ above the galactic plane (0.5, 1, 2, and 3 kpc bottom to top).
  Thin green lines: $K_R$ recovered from a Jeans equation in the case of a thick disc. 
  Thin red lines: the case of a thin disc.
  (Right)
    Relative errors on the recovered radial force $K_R$  versus $R$ at various $z$ (0.5, 1, 2, and 3 kpc) (resp. green, black, red, and blue lines) for thin disc DF (dotted lines) and thick disc DF (continuous lines).    }
    \label{fig6}
\end{figure*}

We tested the efficiency of using the quasi integral $I_3$ to model the distribution function of disc stars using
a Shu distribution function (DF) generalized to 3D axisymmetric potentials. It writes as  \citep[here, we correct a typographic error in][]{Bienayme99}:
\begin{equation}
 \label{eq8}
 \begin{split}
  f(E,L_z, I_3) =
 &\frac{2\Omega(R_c)}{ 2\pi\kappa(R_c)}
 \frac{\Sigma(L_z)}{
            \sigma_R^2 }
             \exp \left[
 -\frac{ \left(E-E_{circ}\right)}{\sigma_R^2}
 \right] 
 \\
 & \times
 \frac{ 1}{ \sqrt{2\pi} } \frac{1}{\sigma_z }
 \exp \left[
 -\left( E-E_{circ} \right)
 \left( \frac{1}{\sigma_z^2}-\frac{1}{\sigma_R^2} \right)  
 I_3 \right] .
 \end{split}
 \end{equation}

\noindent with $R_c(L_z)$  the radius of the circular orbit with the angular momentum $L_z$, 
 $\Omega$  the angular velocity, $\kappa$  the epicyclic frequency, and $E_{circ}$
 the energy of a circular orbiting star at radius $R_c$.  
For sufficiently small velocity dispersions, the number density distribution,
$  \Sigma(L_z) =\Sigma_0 \exp( -R_c/R_{\nu})$,
 is close to $  \Sigma(R) =\Sigma_0 \exp( -R/R_{\nu})$.  
We set constant the parameters $\sigma_{R,z}(L_z)$ and $R_\nu$=2.5\,kpc,
 which is close to the scale length of the number density distribution. 
 The DF allows us to reproduce the triaxiality and tilt of the velocity ellipsoid and to model nearly exponential density disc distribution. It  could be easily modified to reproduce any reasonable radial density law. We set  a constant velocity dispersion ($R_{\sigma_{R,z}}$=$\infty$) for a thin disc ($\sigma_R,\sigma_z$)=(40\,km/s,20\,km/s) and for a thick disc ($\sigma_R,\sigma_z$)=(60\,km/s,40\,km/s).\\

The distribution function can also be written in a more readable form as
 \begin{equation}
 \label{eq9}
 f(E,L_z, I_3) =g(L_z)
             \exp \left[
 -\frac{{\cal E}_R}{\sigma_R^2}
 \right] 
             \exp \left[
 -\frac{{\cal E}_z}{\sigma_z^2}
 \right] 
 ,\end{equation}
where ${\cal E}_R$ and ${\cal E}_z$ are integrals of motion  that can be easily deduced from Eq. \ref{eq8}: 
\begin{equation}
\begin{aligned}
{\cal E}_R= & (E-E_{circ}) \,(1-I_3)
\\
 {\cal E}_z=  &(E-E_{circ}) \,I_3 \, .
\end{aligned}
\end{equation}
They are respectively related to the amount of radial or vertical motions that can be controlled versus the parameters ${\sigma_R}$ and 
${\sigma_z}$. Within a St\"ackel potential and for an orbit with angular momentum $L_z$, they are respectively at position $(R_c(L_z), z=0)$ the radial and vertical kinetic energies. Shell orbits have ${\cal E}_R$=0.
\\

We  determine the moments of the distribution function (Eq. \ref{eq8}) and  recover the radial and vertical  forces, $K_R$ and $K_z$, from the Jeans equations for  a stationary axisymmetric potential.
The moments are linked through the Jeans equations where the time derivative are not exactly zero since  $I_3$ is just an approximate integral:
\begin{equation}
\label{eq10}
\frac{\partial \, \nu \overline{v_R}}{\partial t}+
\frac{\partial}{\partial R}(\nu \overline{v_R^2})
+\frac{\partial}{\partial z}(\nu \overline{v_R v_z})
+\nu \left( \frac{\overline{v_R^2}-\overline{v_\phi^2}}{R}  + \frac{\partial\Phi}{\partial R} \right)
=0 \, ,
\end{equation}
\begin{equation}
\label{eq11}
\frac{\partial \, \nu \overline{v_z}}{\partial t}+
\frac{\partial}{\partial R}(\nu \overline{v_R v_z})
+\frac{\partial}{\partial z}(\nu \overline{v_z^2})
+\nu \left( \ \frac{\overline{v_R v_z}}{R}  + \frac{\partial \Phi}{\partial z} \right)
=  0  \, .
\end{equation}
By considering that $I_3$ is close to an exact integral and  neglecting the time derivative terms, we transform the  Jeans equations  as
\begin{equation}
\label{eq12}
\frac{\partial}{\partial R}(\nu \overline{v_R^2})
+\frac{\partial}{\partial z}(\nu \overline{v_R v_z})
+\nu \left( \frac{\overline{v_R^2}-\overline{v_\phi^2}}{R}  + \left( \frac{\partial\Phi}{\partial R} \right)_{est.} \right)
=0 \, ,
\end{equation}
\begin{equation}
\label{eq13}
\frac{\partial}{\partial R}(\nu \overline{v_R v_z})
+\frac{\partial}{\partial z}(\nu \overline{v_z^2})
+\nu \left( \ \frac{\overline{v_R v_z}}{R}  + \left( \frac{\partial \Phi}{\partial z} 
  \right)_{est.}  \right)
=  0 \, .
\end{equation}

Thus, under the assumption that the approximate integral $I_3$  is  exact, Eqs. \ref{eq12}-\ref{eq13} are used to obtain an estimate of the radial and vertical forces, 
 $K_R \approx - \left( \frac{\partial \phi}{\partial R} \right)_{est.}$
  $K_z \approx - \left(\frac{\partial \phi}{\partial z} \right)_{est.}$.
They can be compared to the exact forces
to test  the efficiency to recover the galactic potential using our St\"ackel integral in a non-St\"ackel potential.
Figure \ref{fig6} shows the radial forces $K_R$ of the BGM potential and its relative error at several $z$ when recovered from  Eqs \ref{eq12}-\ref{eq13}.  

Within the domain $R$\,=\,3 to 16\,kc and $z$ smaller than 3\,kpc, the relative error on the  radial force $K_R$ is smaller than one per cent for $R>6$\,kpc. The relative error  is ten  per cent at $R$\,=\,3\,kpc and increases  below $R$\,=\,3\,kpc.

\begin{figure}[!htbp]
\begin{center}
\resizebox{8.5cm}{!}{\rotatebox{-90}{\includegraphics{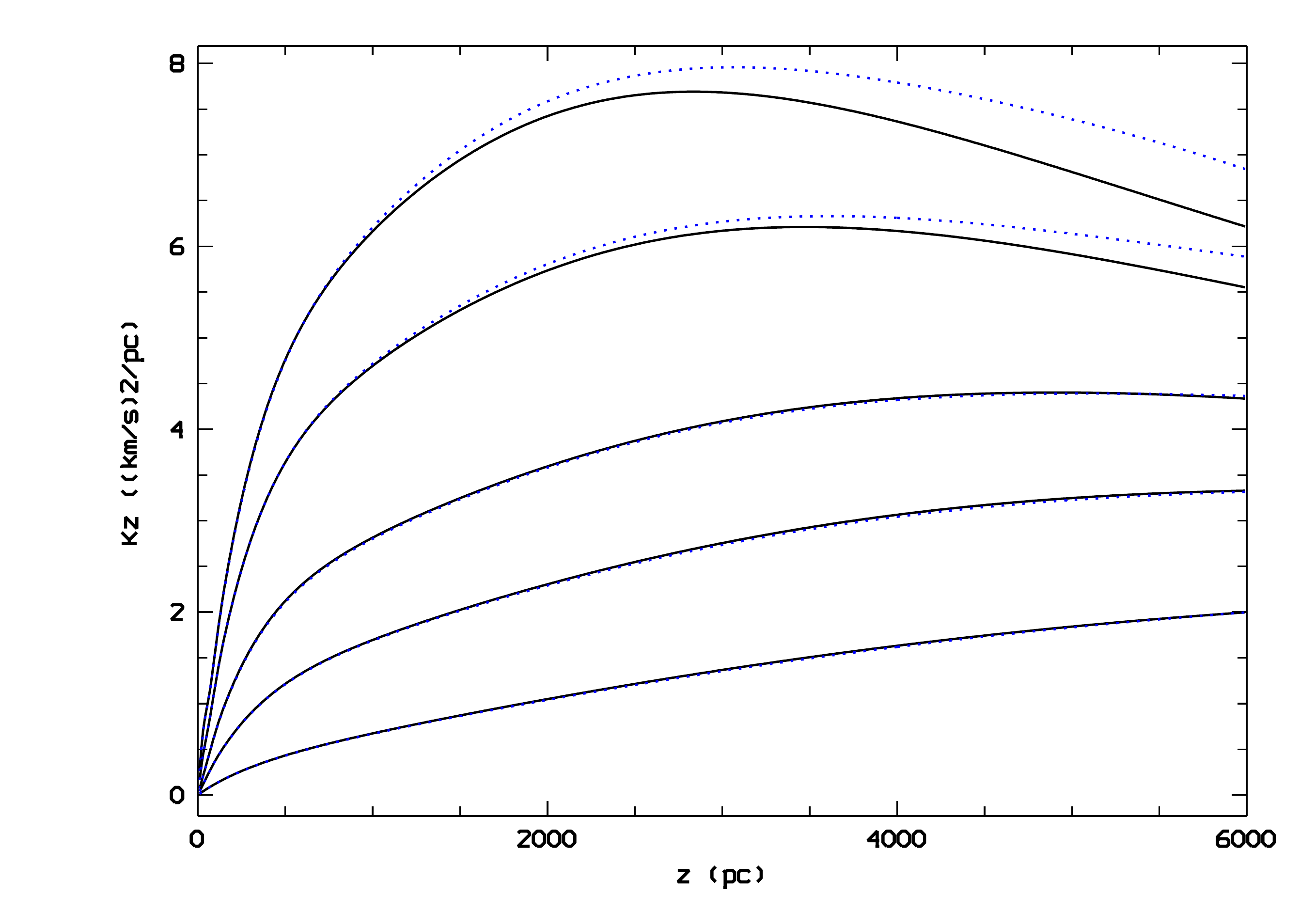}}}
\end{center}
  \caption{ Vertical force $K_z$  versus $z$ at  $R$=(3.5,4.5,6.5,8.5,12.5 kpc) (top to bottom) from the BGM  (black lines) and recovered $K_z$ for a thick disc DF (blue dotted lines).    }
    \label{fig7}
\end{figure}

\begin{figure}[!htbp]
\begin{center}
\resizebox{8.5cm}{!}{\rotatebox{-90}{\includegraphics{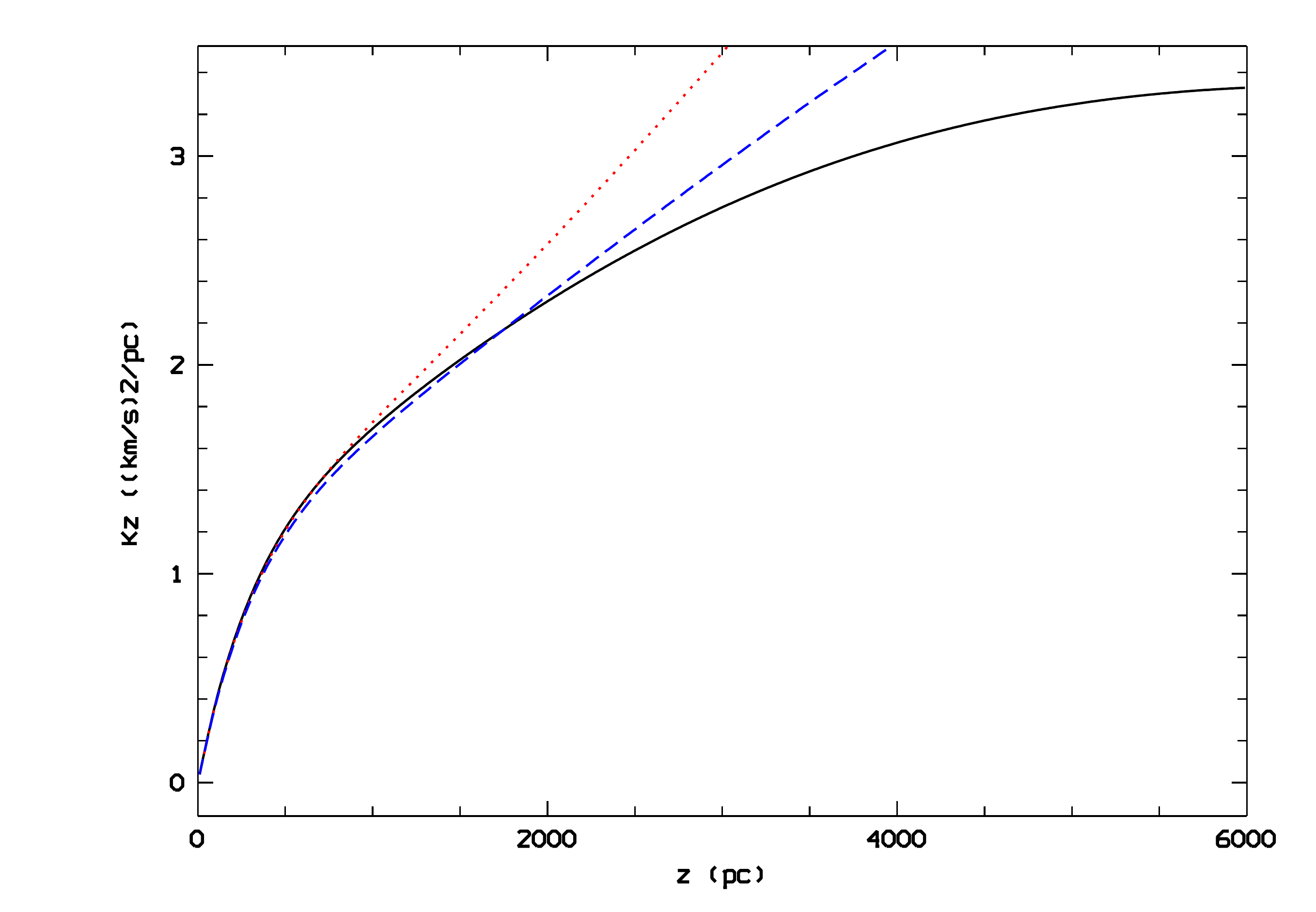}}}
\end{center}
  \caption{ Top : Vertical force $K_z$  versus $z$ at  $R$=8500 pc (black line) and recovered $K_z$ from simplified Jeans equation assuming  the separability of vertical and radial motions:  red dotted line for a thin disc DF and blue dashed lines for a thick disc DF.
    }
    \label{fig8}
\end{figure}

Figure \ref{fig7} shows the exact and recovered $K_z$ force versus $z$ at several $R$ (3.5, 4.5, 6.5, 8.5, 12.5). At large $R >$ 6\,kpc,  the $K_z$ forces are accurately recovered for the thick disc up to 6\,kpc and up to 3\,kpc for the
thin disc.
The vertical force  $K_z$ at the solar position $R$=8\,kpc is remarkably well recovered at 0.5\% up to 6\,kpc for the thick disc (7 scale heights) (Figure \ref{fig8}). For the thin disc, the accuracy is lost above 3 kpc (corresponding however to ten scale heights of the thin disc).
At lower radius $R$ from 3.5 to 5, the $K_z$ force is only approximately recovered up to $z$=2\,kpc.

For comparison (Figure \ref{fig8}), we also plot the $K_z$ force recovered neglecting the cross term $<v_R v_z>$ in Jeans equations, a classical assumption valid at low $z$. We note that this hypothesis remains valid up to 500 pc for the thin disc. At higher $z$, the recovered force diverges quickly from the exact one.
%
%
%
\section{Collisionless Boltzmann equation}
The stationarity of the Jeans equation is a necessary condition for validating the degree of stationarity of a distribution function, but  this is not a sufficient condition. We should also have to test the stationarity of all the other moments of the collisionless Boltzmann equation (CBE), since  it is easy to find solutions for the Jeans equations that are not solutions of the CBE, so the Jeans equation test can be  more optimistic than  the indications solely obtained from the conservation of the integral of motion. For this reason we examine hereafter the stationarity of the CBE.\\

\subsection{First estimate}

We  estimate the stationarity of the distribution function built with the approximated third integral by looking at the time variation of the DF (Eq. \ref{eq8}) over a dynamical time ($\sim$ an orbit revolution).
We have
\begin{equation}
\label{eq14}
\frac{\mathrm d \ln f}{\mathrm d t}
=
\frac{\partial \ln f}{\partial t}+\left[ \ln f, H \right]
=
0
\end{equation}
or
\begin{equation}
\label{eq15}
\frac{\partial \ln f}{\partial t}
-   \frac{\Delta E}{\sigma_R^2}  \left(  \frac{\sigma_R^2}{\sigma_z^2}- 1  \right) \frac{\mathrm d I_3}{\mathrm d t} =0
,\end{equation}
and the relative variation of $f$ over a dynamical or  longer time is determined by the variation $\sigma_{I3}$ of the quasi integral $I_3$ along orbits:
\begin{equation}
\label{eq16}
 \left|  \frac{\Delta f}{f} \right| 
  \sim    \frac{\Delta E}{\sigma^2}  \, \sigma_{I3} \, .
\end{equation}
Orbits with small radial or vertical amplitudes ($\Delta E$ and $\sigma_{I3}$ small) have the smallest  variations in $f$.
Thus the thinnest discs are the most accurately modelled.

The stationarity decreases quickly with $z$  since we have $\Delta E/\sigma_R^2 \propto  z^{\sim0.5}$. It also decreases because of the significant presence of resonant orbits between 1 to 2\,kpc.
However, at the solar Galactic radius $R_0$, orbits within the BGM with large vertical amplitude remain correctly modelled with a St\"ackel potential thanks to small  $\sigma_{I3}$ of the order of 0.005.
In this situation, the distribution function remains stationary at  many scale heights above the galactic plane. For the thick disc, the stationarity (Eq. \ref{eq16}) is about one per cent at six scale heights ($z \sim$ 6\,kpc). For the thin disc, it becomes insufficient at about ten scale heights ( $z \sim$ 3\,kpc). 

\begin{figure}[!htbp]
\begin{center}
\resizebox{8.5cm}{!}{\rotatebox{-90}{\includegraphics{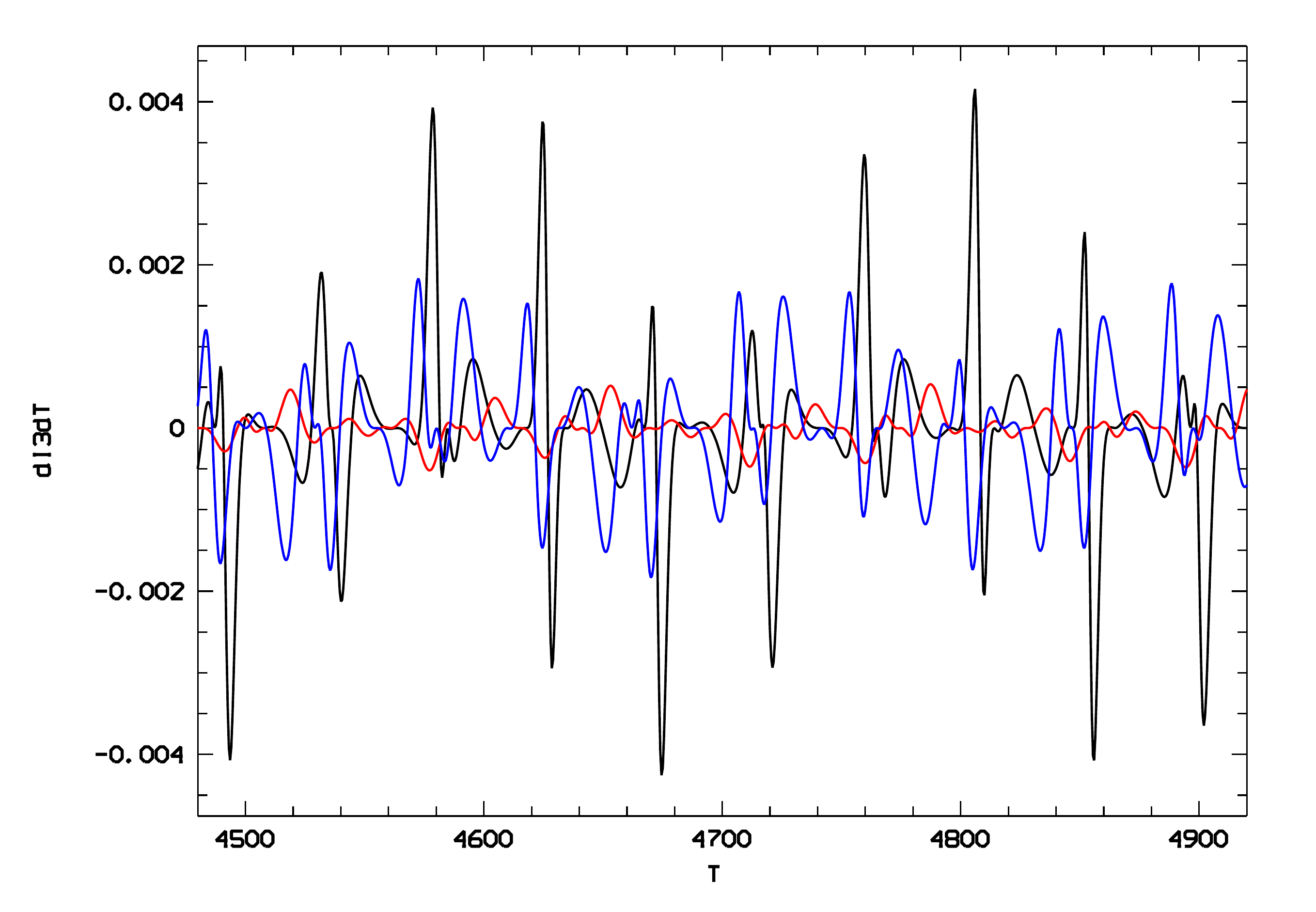}}}
\end{center}
  \caption{ ${\rm d} I_3 / {\rm d} t$: time derivative of the  quasi integral within the logarithmic potential for the three orbits shown in Fig \ref{fig1}: (black $I_3=0.16$, red $I_3=0.88$, blue $I_3=1.06$).
    }
    \label{fig9}
\end{figure}

The time derivative in Eq \ref{eq14} is the time variation from an initial position of stars in phase space at $t$=0 given by the initial condition of Eq. \ref{eq9}. How divergent are the orbits evolving from this supposedly near-equilibrium initial condition? We know that for potentials similar to the three  potentials considered in this paper, the orbits are essentially regular and an effective third integral must exist, which could be eventually estimated with a higher accuracy using, for instance,  high-order polynomials   \citep{Bienayme13} or a torus fitting \citep{Sanders14}. This would, however, not be true in  cases of significant presence of ergodicity, for instance close to the corotation of a barred rotating potential.
It turns out that our approximate integrals oscillate along each orbit around a mean value,  that a typical periodicity of these oscillations is the dynamical time, and that the amplitude of these oscillations is of the order of $\sigma_{I3}$. This is illustrated in Fig. \ref{fig9} where  the maxima of ${\rm d} I_3 / {\rm d} t$ are of the order of $\sigma_{I3}/t_{dyn}$ with $t_{dyn}\sim6$.
%
%

\subsection{Second estimate}

It is also useful to consider the non-stationarity as a shift in positions or velocities of the modelled DF relative to a stationary DF.
A simple 1D dimensional analysis allows it to be illustrated.
With a quadratic potential $\Phi=a z^2$ and a stationary DF as
\begin{equation}
 f_{stat.}=\exp \left( -(\phi(z)+v_z^2/2)/\sigma^2 \right) \, ,
\end{equation}
the shifted DF in velocity by a factor $v_0$ is
\begin{equation}
f=\exp \left[ -\frac{1}{\sigma^2} \left(\phi(z)+\frac{(v_z-v_0)^2}{2} \right) \right] \, .
\end{equation}
Such a DF could be considered satisfying if the shift is small even if we notice that   the relative errors on $f$ increase  in the tails of the distribution.\\

We deduce from the CBE,
\begin{equation}
\frac{\partial   \ln f}{ \partial  t}= -\frac{K_z}{\sigma^2 } \,v_0
,\end{equation}
which combined with Eq. \ref{eq16}, gives the variation in $f$ over a dynamical time $t_{dyn}$:
\begin{equation}
\frac{|K_z|}{\sigma^2 } \,{v_0}
= \frac{\Delta E}{\sigma^2}  \, \frac{\sigma_{I_3}}{t_{dyn}} \, .
\end{equation}
With a quadratic potential and  $h$, the scale height of the disc population, we estimate the velocity shift:
\begin{equation}
v_0
\sim\frac{\Delta E}{|K_z|}  \, \frac{\sigma_{I3}}{t_{dyn} }
\sim     \frac{z}{h } \sigma \, \sigma_{I3}  \, .
\end{equation}
For the thin disc DF at one scale height ($h \sim$ 250\, pc),  the shift $v_0$ is  0.2 km/s  and at 4 kpc (16 scale heights), it is 10 km/s (to be compared to the 20 km/s of the vertical velocity dispersions) where the application of the Jeans equation shows that the DF is not stationary.
At one scale height of $\sim$ 1\,kpc for the thick disc, the shift is $v_0\sim$ 1 km/s. At 6 kpc, it is 7 km/s (to be compared to the 40 km/s of the vertical velocity dispersion), and the Jeans equation still validates the stationarity.

\section{Conclusion}

In this paper we  have shown that St\"ackel potentials can be used to fit a wide variety of disc stellar orbits within different axisymmetric potentials of disc galaxies.
We also proposed  a new and simple formulation for the third integral of motion of Galactic potentials, explicitly depending on the potential, which is an integral  known to be exact in the case of St\"ackel potentials.

The quality of the fit of orbits was quantitatively measured by looking at the conservation of the approximate third integral of motion besides the energy and angular momentum. By using the Besan\c con Galactic model, we showed that the third integral is conserved to a few  thousandths in a wide volume around the solar neigbourhood. It is conserved to better than one per cent at 6 kpc above the Galactic plane at the solar position. However, the fit fails at low galactic radius ($R$ smaller than 4 kpc) owing to the presence of a large number of resonant orbits.

In light of the future Gaia data, we also used distribution functions of disc stars, depending on the three integrals, and checked the ability of Jeans equation to recover the gravitational potential. We also considered the stationarity of the CBE.

In conclusion, St\"ackel potentials can be  good local approximations of general realistic potentials. They have  been used many times to fit the global potential of galaxies 
\citep[see for instance][]{Dejonghe88} or the potential of our own Galaxy  \citep{Famaey03}.
The fit of stellar orbits are also frequently used using St\"ackel potentials with local or global fitting \citep{Kent91}.  We showed here that they be can used very efficiently to build distribution function of disc stellar populations. A straightforward application will consist in using such modelling of DFs and in extending the dynamical consistency of the Besan\c con Galactic model to describe the kinematics of disc stellar populations.

%

\bibliographystyle{aa} 
\bibliography{} 

\begin{thebibliography}{56}
\expandafter\ifx\csname natexlab\endcsname\relax\def\natexlab#1{#1}\fi


\bibitem[Batsleer 
\& Dejonghe(1994)]{Batsleer94} Batsleer, P., \& Dejonghe, H.\ 1994, \aap, 287, 43 


\bibitem[Bienaym{\'e}(1999)]{Bienayme99} Bienaym{\'e}, O.\ 1999, \aap, 341, 86 

\bibitem[Bienaym{\'e}(2009)]{Bienayme09} Bienaym{\'e}, O.\ 2009, \aap, 500, 781 


\bibitem[Bienaym{\'e} et 
al.(2014)]{Bienayme14} Bienaym{\'e}, O., Famaey, B., Siebert, A., et al.\ 2014, \aap, 571, A92 

\bibitem[Bienaym\'e et 
al.(1987)]{Bienayme87} Bienaym\'e, O., Robin, A.~C., \& Cr\'ez\'e, M.\ 1987, \aap, 186, 359

\bibitem[Bienaym{\'e} 
\& Traven(2013)]{Bienayme13} Bienaym{\'e}, O., \& Traven, G.\ 2013, \aap, 549, A89 

\bibitem[Binney(2012)]{Binney12} Binney, J.\ 2012, \mnras, 426, 
1324 

\bibitem[Colombi et al.(2015)]{Colombi15} Colombi, S., Sousbie, 
T., Peirani, S., Plum, G., \& Suto, Y.\ 2015, arXiv:1504.07337, MNRAS in press 

\bibitem[Czekaj et 
al.(2014)]{Czekaj14} Czekaj, M.~A., Robin, A.~C., Figueras, F., Luri, X., \& Haywood, M.\ 2014, \aap, 564, A102 

\bibitem[De Bruyne et al.(2000)]{deBruyne00} De Bruyne, V., 
Leeuwin, F., \& Dejonghe, H.\ 2000, \mnras, 311, 297 

\bibitem[Dejonghe 
\& de Zeeuw(1988)]{Dejonghe88} Dejonghe, H., \& de Zeeuw, T.\ 1988, \apj, 333, 90 

\bibitem[de Zeeuw(1985)]{deZeeuw85a} de Zeeuw, T.\ 1985, \mnras, 
216, 273 

\bibitem[de Zeeuw 
\& Lynden-Bell(1985)]{deZeeuw85b} de Zeeuw, P.~T., \& Lynden-Bell, D.\ 1985, \mnras, 215, 713 

\bibitem[Famaey 
\& Dejonghe(2003)]{Famaey03} Famaey, B., \& Dejonghe, H.\ 2003, \mnras, 340, 752 

\bibitem[Fehlberg (1968)]{Fehlberg68} Fehlberg, E.\ 1968,  
 NASA technical report  TR R-287          


\bibitem[Hori(1962)]{Hori62} Hori, G.\ 1962, \pasj, 14, 353 

\bibitem[Kent 
\& de Zeeuw(1991)]{Kent91} Kent, S.~M., \& de Zeeuw, T.\ 1991, \aj, 102, 1994 

\bibitem[Manabe(1979)]{Manabe79} Manabe, S.\ 1979, \pasj, 31, 
369 

\bibitem[Lynden-Bell(1962)]{Lynden-Bell62} Lynden-Bell, D.\ 1962, 
\mnras, 124, 95


\bibitem[Ollongren(1962)]{Ollongren62} Ollongren, A.\ 1962, \bain, 
16, 241 

\bibitem[Papaphilippou 
\& Laskar(1998)]{Papaphilippou98} Papaphilippou, Y., \& Laskar, J.\ 1998, \aap, 329, 451

\bibitem[Piffl et al.(2014)]{Piffl14} Piffl, T., Binney, J., 
McMillan, P.~J., et al.\ 2014, \mnras, 445, 3133 

\bibitem[Renaud et al.(2013)]{Renaud13} Renaud, F., Bournaud, 
F., Emsellem, E., et al.\ 2013, \mnras, 436, 1836

\bibitem[Richstone(1980)]{Richstone80} Richstone, D.~O.\ 1980, 
\apj, 238, 103 

\bibitem[Robin et 
al.(2003)]{Robin03} Robin, A.~C., Reyl{\'e}, C., Derri{\`e}re, S., \& Picaud, S.\ 2003, \aap, 409, 523 

\bibitem[Sanders 
\& Binney(2014)]{Sanders14} Sanders, J.~L., \& Binney, J.\ 2014, \mnras, 441, 3284 


\bibitem[Syer 
\& Tremaine(1996)]{Syer96} Syer, D., \& Tremaine, S.\ 1996, \mnras, 282, 223 

\bibitem[Valluri 
\& Merritt(1998)]{Valluri98} Valluri, M., \& Merritt, D.\ 1998, \apj, 506, 686 

\bibitem[van de Hulst(1962)]{vandeHulst62} van de Hulst, H.~C.\ 
1962, \bain, 16, 235 

\bibitem[Wayman(1959)]{Wayman59} Wayman, P.~A.\ 1959, \mnras, 
119, 34 

\bibitem[Yoshikawa et al.(2013)]{Yoshikawa13} Yoshikawa, K., 
Yoshida, N., \& Umemura, M.\ 2013, \apj, 762, 116 

\bibitem[Zotos(2011)]{Zotos11} Zotos, E.~E.\ 2011, \na, 16, 391 

\end{thebibliography}

\appendix

\section{Quasi integral of motion}

We refer the reader to  \cite{deZeeuw85a} and \cite{deZeeuw85b} for notations and for a detailed description of properties of St\"ackel potentials.

An axisymmetric St\"ackel  potential is fully defined by two free functions $h(\lambda)$ and $h(\nu)$. In the case of prolate spheroidal coordinates, $\pm z_0$ ($z_0^2=\gamma-\alpha$ ) are   the foci  of the confocal ellipsoids and hyperboloids used to define  a  system  of coordinates $(\lambda,\nu)$ in which St\"ackel potentials are more easily tractable.

Let $V(R,z)=\tilde V(\lambda,\nu)$  be the true potential that we approximate locally  at the position $(R_1,z_1)= (\lambda_1,\nu_1)$ with a St\"ackel potential $\Phi(\lambda,\nu)$. We  assume that $z_0$ is already known.

By definition, we have for a St\"ackel potential:
\begin{equation}
\Phi(\lambda,\nu)=-\frac{h(\lambda)-h(\nu)}{\lambda-\nu} \, ,\end{equation}

\noindent and it is always possible to find a  S\"ackel potential that coincides with any potential
on the chosen coordinate  surfaces  $\lambda=\lambda_1$ and $\nu=\nu_1$.
It writes as
\begin{equation}
\label{eqA2}
\begin{split}
{ \Phi(\lambda,\nu)=  }\\
&\frac{\tilde V(\lambda,\nu_1)(\lambda-\nu_1)
- \tilde V(\lambda_1,\nu_1)(\lambda_1-\nu_1)
+\tilde V(\lambda_1,\nu)(\lambda_1-\nu)}
{\lambda-\nu}  \, .
\end{split}
\end{equation}
Thus
\begin{equation}
\begin{split}
h(\lambda)= & 
\tilde V(\lambda,\nu_1)(\lambda-\nu_1)
- \tilde V(\lambda_1,\nu_1)(\lambda_1-\nu_1)+ C
\\ 
h(\nu) = &
-\tilde V(\lambda_1,\nu)(\lambda_1-\nu) + C
\end{split}
\end{equation}
and $C$  an arbitrary constant.\\

We can express the 
third integral $I_s$ associated to $\Phi$ as
\begin{equation}
\label{eqA5}
I_s= \Psi(\lambda,\nu)
-\frac{1}{2} \frac{z^2}{z_0^2} (v_R^2+v_\theta^2)
-\frac{1}{2}\,   \left (1+\frac{R^2}{z_0^2}\right)\, v_z^2
+\frac{R\, z\, v_R\, v_z}{z_0^2}\, ,
\end{equation}
or
\begin{equation}
\label{eqA6}
I_s= \Psi(\lambda,\nu)
-\frac{1}{2} \frac{L^2-L_z^2}{z_0^2} -\frac{1}{2} v_z^2
\end{equation}
%

 %
\noindent with 
\begin{equation}
\Psi(\lambda,\nu)=\frac{(\nu+\gamma)\, h(\lambda) -(\lambda+\gamma)\,h(\nu)}
{(\gamma-\alpha)\,(\lambda-\nu)}\, .
\end{equation}
We fix the remaining free constant $C$ by setting $h(\nu=-\gamma)=0$, so $\Psi$ is null at $z=0$ in the plane of symmetry of the potential. Then, evaluated at $(\lambda_1,\nu_1)$, this function 
simplifies as
\begin{equation}
\label{eqA8}
\Psi(\lambda_1,\nu_1)=-\frac{
\left(\tilde V (\lambda_1,\nu_1) 
-  \tilde V(\lambda_1,-\gamma )\right) \,(\lambda_1 +\gamma)}
{\gamma-\alpha}\, .
\end{equation}

If we set $\gamma=z_0^2$ and $\alpha=0,$ 
\begin{eqnarray}{}
\tilde V(\lambda_1,\nu_1)  = & V(R_1,z_1) \,   ,\\
\tilde V(\lambda_1,-\gamma )  = & V(R=\sqrt{\lambda_1},z=0) \, ,
\end{eqnarray}
and
\begin{eqnarray}
 \lambda_1  = & \frac{1}{2}(R_1^2+z_1^2-z_0^2)
+\frac{1}{2}\sqrt{(R_1^2+z_1^2-z_0^2)^2+4R_1^2z_0^2} \, .
\end{eqnarray}
Thus, from Eqs. \ref{eqA5} and \ref{eqA8} (similar to Eq. \ref{eq2}), we obtain a   simple expression for the third integral $I_s$ at position $(\lambda_1,\nu_1 )$, which is exact in the case of a St\"ackel potential and explicitly depends  on the potential, the coordinates, and the velocities.  We note  that, in practice,  the  intermediate functions $h(\lambda)$ and $h(\nu)$  do not need to be evaluated.

In summary, the proposed quasi integral in Eq.~\ref{eq1} results from the exact third integral of the orbits in the St\"ackel potential of Eq.~\ref{eqA2}, which is equal to the true potential on the surfaces $\lambda=\lambda_1$ and $\nu=\nu_1$. The integral is re-evaluated locally for each point $(\lambda,\nu)=(\lambda_1, \nu_1)$.

The case of St\"ackel potentials with oblate spheroidal coordinates leads to the same equations, but with $z_0$ imaginary, $z_0^2<0$ and $\gamma < \alpha$.
In the $z_0=\infty$ limit, 
a St\"ackel potential  is separable in $R$ and $z$,
$$V(R,z)=V_1(R)+V_2(z) \, ,$$
and
$$ -I_s= \left[ V_2(z)-V_2(0) \right] +v_z^2/2 \, .$$
Thus the vertical energy is an integral of motion.
\\

In the  $z_0=0$ limit, St\"ackel potentials have the form
$$V(R,z)=V_1(r)+\frac{V_2(\theta)}{r^2} ,$$
where  $r^2=R^2+z^2$ and $\theta$ is the polar angle \citep[][Table 1]{Lynden-Bell62}, and  the integral of motion is
$$-\,z_0^2 I_s   \rightarrow   [V_2(\theta)-V_2(\pi/2)] +\frac{1}{2}(L^2-L_z^2)\, .$$

\end{document}